# Intrinsic signal optoretinography of dark adaptation abnormality due to rod degeneration

**Jie Ding[a], Tae-Hoon Kim[a], Guangying Ma[a], Xincheng Yao[a,b*]**

[a] University of Illinois Chicago, Department of Biomedical Engineering, Chicago, IL 60607, USA
[b] University of Illinois Chicago, Department of Ophthalmology and Visual Sciences, Chicago, IL 60607, USA

**Abstract**

**Significance:** Multiple eye diseases such as age-related macular degeneration, diabetic retinopathy, and retinitis pigmentosa can cause photoreceptor dysfunction. Rod photoreceptors are known to be more vulnerable than cone photoreceptors. Therefore, functional assessment of rod photoreceptors is important for early detection of eye diseases.

**Aim:** This study is to demonstrate the feasibility of using intrinsic optical signal (IOS) optoretinography (ORG) for objective detection of dark adaptation (DA) abnormality due to rod photoreceptor degeneration.

**Approach:** Functional optical coherence tomography (OCT) was employed for IOS ORG of wild-type (WT) and retinal degeneration 10 (rd10) mice. Six WT C57BL/6J and eight rd10 B6.CXB1-Pde6brd10/J mice at postnatal day 14 (P14) were used for this study. Dynamic OCT analysis of retinal thickness and brightness changes, corresponding to light to dark transition, was implemented

**Results:** Comparative measurement of the retina under light and dark conditions revealed significant IOS changes within the outer retina. The thickness between external limiting membrane (ELM) and retinal pigment epithelium (RPE) reduced, and the OCT brightness of inner segment ellipsoid zone (EZ) decreased during DA, compared to light adaptation (LA). Relative EZ intensity change was observed to decrease larger in rd10, compared to WT. The relative intensity of the hyporeflective band between ELM and RPE also showed significant decrease in rd10 retinas during DA.

**Conclusions:** DA-IOS abnormalities were observed in rd10, compared to WT at P14. The ORG measurement of DA-IOS kinetics promises valuable information for noninvasive assessment of photoreceptor function.

*Corresponding author, E-mail: xcy@uic.edu

# 1 Introduction

Eye diseases, such as age-related macular degeneration (AMD)[1-3], retinitis pigmentosa (RP)[4,5], and diabetic retinopathy (DR)[6,7], can cause photoreceptor dysfunctions. Particularly, retinal photoreceptors are known as the primary target of both AMD and RP. Currently, there is no outright cure for the retinal degenerative disease that produces irreversible damage to photoreceptors and retinal pigment epithelium (RPE). A key strategy for preventing vision loss due to retinal degeneration is to preserve function and be vigilant for changes in signifying progression. Therefore, early detection of potential biomarkers is essential to enable prompt treatment in preventing or slowing retinal degenerative diseases[8,9]. Structural biomarkers, such as drusen and pigmentary abnormalities in the macula of AMD patients, have provided valuable information for evaluating eye conditions. However, the morphological examination is often not sufficient[10]. In principle, physiological abnormality in diseased cells can occur before detectable morphological changes in the retina, such as retinal cell loss and corresponding thickness change. Therefore, functional evaluation of physiological integrity of retinal cells is desirable for the early detection of retinal diseases.

Intrinsic optical signal (IOS) imaging[11-19], also termed as optophysiology[20], optoretinogram[21-25] or optoretinography (ORG)[15,26-30], is based on dynamic measurement of near-infrared light response in the retina activated by visible light stimulation. So, concurrent assessment of retinal morphology and physiology is readily available. Since optical coherence tomography (OCT) is a non-invasive method to enable depth-resolved imaging of individual retina layers, OCT-based intrinsic signal ORG has been actively explored in the human and animal retinas[15,19,21-36].

Other approaches such as electroretinography (ERG) have revealed dark adaptation (DA) abnormality as an early symptom of rod photoreceptor degeneration[37,38]. Comparative OCT of the retina in light adaptation (LA) and DA has also revealed outer retina changes[39-41]. Recently, Kim et al. has demonstrated intrinsic signal ORG monitoring of DA kinetics, including shortening of the outer retina, a rearrangement of the photoreceptor interdigitation zone, and a reduction in intrinsic signal amplitude at the ellipsoid zone (EZ), in the healthy mouse retina[42]. However, dynamic monitoring of structural and physiological changes in the diseased retina remains to be explored, which is crucial to validate the feasibility of ORG assessment for DA abnormality.

The present study aims to demonstrate DA-IOS abnormality in retinal degeneration 10 (rd10) mouse model. Rd 10 is a well-established mouse model for the study of rod photoreceptor degeneration[43-47]. In rd10 mice, a spontaneous mutation of the β-subunit of rod-phosphodiesterase (PDE) gene leads to the accumulation of cGMP and then produces rod photoreceptor degeneration[46-51]. As the rods still express about 40% of endogenous PDE in the rd10 mouse, the rd10 manifests a relatively slow onset of rod photoreceptor degeneration at around postnatal day 17 (P17)[50,52], which makes it a better mouse model to study rod degeneration. Thus, we measured DA-IOS in rd10 mice at P14, when the retina is morphologically intact. We analyzed DA-induced morphological and physiological change in both wild-type (WT) and rd10 mouse retinas, therefore, to validate the application of using functional OCT for objective ORG measurement of DA-IOS kinetics.

## 2 Material and Method

*2.1 Animal*

Six C57BL/6J (WT) and eight B6.CXB1-Pde6brd10/J (rd10) mice (Jackson Laboratory, Bar Harbor, Maine, USA) at postnatal day 14 (P14) were used in this study. Before the DA experiment, mouse eyes were light adapted for more than 5 hours[40]. Mice were anesthetized with a mixture of 100 mg/kg ketamine and 5 mg/kg xylazine, by intraperitoneal injection. After anesthesia, mouse eyes were dilated with 0.5% tropicamide. Lubricant eye gel (GenTeal, Alcon Laboratories, Fort Worth, Texas, USA) was used during experiments to prevent eye dryness. After the experiment, mice were awake and housed in the animal room of the Biological Resources Laboratory building at the University of Illinois Chicago. All animal handling procedures were conformed to the Association for Research in Vision and Ophthalmology statement for the use of animals in ophthalmic and vision research. All experiments were performed following the protocols approved by the Animal Care Committee at the University of Illinois Chicago.

## 2.2 Imaging System

A custom-designed OCT system was employed for *in vivo* ORG measurement of the mouse retina[42]. A superluminescent diode (SLD) was used as the light source ($\lambda$=810 nm, $\Delta\lambda$=100 nm, D-810-HP, Superlum, Carrigtwohill, County Cork, Ireland). A fiber coupler (75:25, TW850R5A2, Thorlabs, Newton, New Jersey, USA) divided the light to the sample and reference arms. A custom-designed spectrometer was constructed with a line CCD camera (2048 pixels, OCTOPLUS, e2v, Chelmsford, UK) and a transmission grating (1200 line/mm, Wasatch Photonics, West Logan, Utah, USA). The axial and lateral resolution of the system was theoretically estimated at 2.9 μm and 11 μm, respectively.

## 2.3 Data Acquisition and Processing

Time-lapse OCT data were collected in a laboratory room under ambient light condition during the early afternoon to minimize the effect of disk shedding caused by circadian rhythm[40,53]. For a total 30 min recording, 7 OCT volumes were sequentially acquired with a 5 min interval. OCT scanning area was selected at the dorsal quadrant (Fig. 1a). Each OCT volume consisted of 500 B-scans, and each B-scan contained 500 A-lines, over a 1.4 × 1.4 mm area. The imaging speed was at 80 B-scans per second. The first OCT was recorded under ambient light condition, and the other 6 OCT volumes were sequentially recorded under dark condition.

For image processing, each OCT volume was reconstructed on MATLAB. The workflow was k-sampling, interferogram extraction, apodization, fast Fourier transform (FFT), and image registration. After OCT reconstruction, 40 adjacent B-scans at the center of OCT volume were averaged into one B-scan (Fig. 1b). The averaged B-scan was flattened by realigning A-lines. Then, a center region of the flattened B-scan spanning 80-100 A-lines was cut and averaged to produce a single A-line reflectance profile. Next, the intensity of the A-line profile was normalized based on the intensity of outer nuclear layer (ONL).

*2.4 Statistical Analysis*

For statistical analysis, retinal thickness and intensity of individual layers were measured based on the normalized A-line profile. Statistical analysis was done by using one-way repeated-measures analysis of variance (ANOVA) with Bonferroni correction using Origin. The *p*-values $< 0.05$ were considered as significant. Error bar was added in each datapoint as standard deviation.

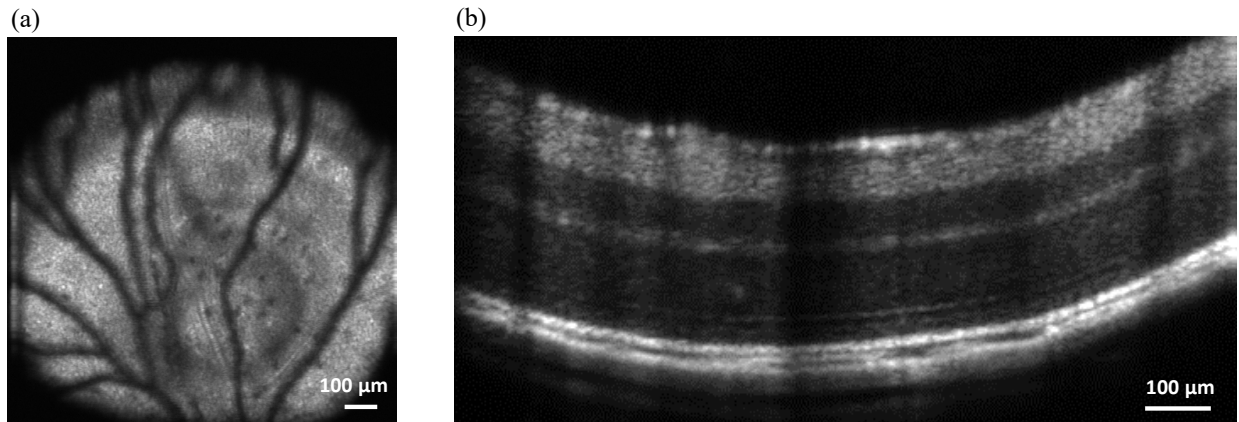

**Fig. 1** (a) en face projection image of the dorsal quadrant in the mouse retina. (b) Representative averaged OCT B-scan of the mouse retina. Scale bar: 100 μm.

## 3 Result

### 3.1 Comparative Imaging of The Retina with Light and Dark Adaptations

Comparative OCT of WT and rd10 mice with LA and DA was implemented at P14. As shown in Fig. 2a, the inner retina of both WT (Fig. 2a1) and rd10 (Fig. 2a2) mice did not show a notable difference in LA and DA condition. However, the region between the external limiting membrane (ELM) and RPE (red arrows, Fig. 2a) showed a thickness reduction in DA condition. Thus, we further analyzed DA-induced changes in the outer retina.

In Fig. 2b, the averaged A-line profiles of the outer retina from six WT and eight rd10 mice retinas were illustrated to verify the outer retina shrinkage. For easy comparison, the ELM band peaks were set to 0 on the X-axis. As shown in Fig. 2b, it was observed that RPE bands shift towards the ELM in DA condition, compared to that in LA condition. This observation consistently supports that the outer retina thickness, i.e., the distance between the OPL and RPE bands, is reduced with DA.

In addition to the outer retinal shrinkage, OCT band brightness was also altered. The OPL, ELM, and RPE peak values did not show significant differences in LA and DA conditions. However, EZ peak intensity was notably dropped in DA condition, compared to that in LA condition (Fig. 2b).

In addition to the hyper-reflective bands at OPL, ELM, EZ, and RPE, two hypo-reflective bands T1 (between ELM and EZ) and T2 (between EZ and RPE) were clearly observed in WT retinas under both LA and DA conditions (Fig. 2b1). On the contrary, T2 was elusive in rd10, while T1 was rather clear (Fig. 2b2). Since the T2 is located between the EZ and RPE, it mostly corresponds to the photoreceptor outer segment (OS). This observation suggests possible photoreceptor OS abnormality, i.e., physiological defect of the photoreceptor, in rd10, compared to WT at P14.

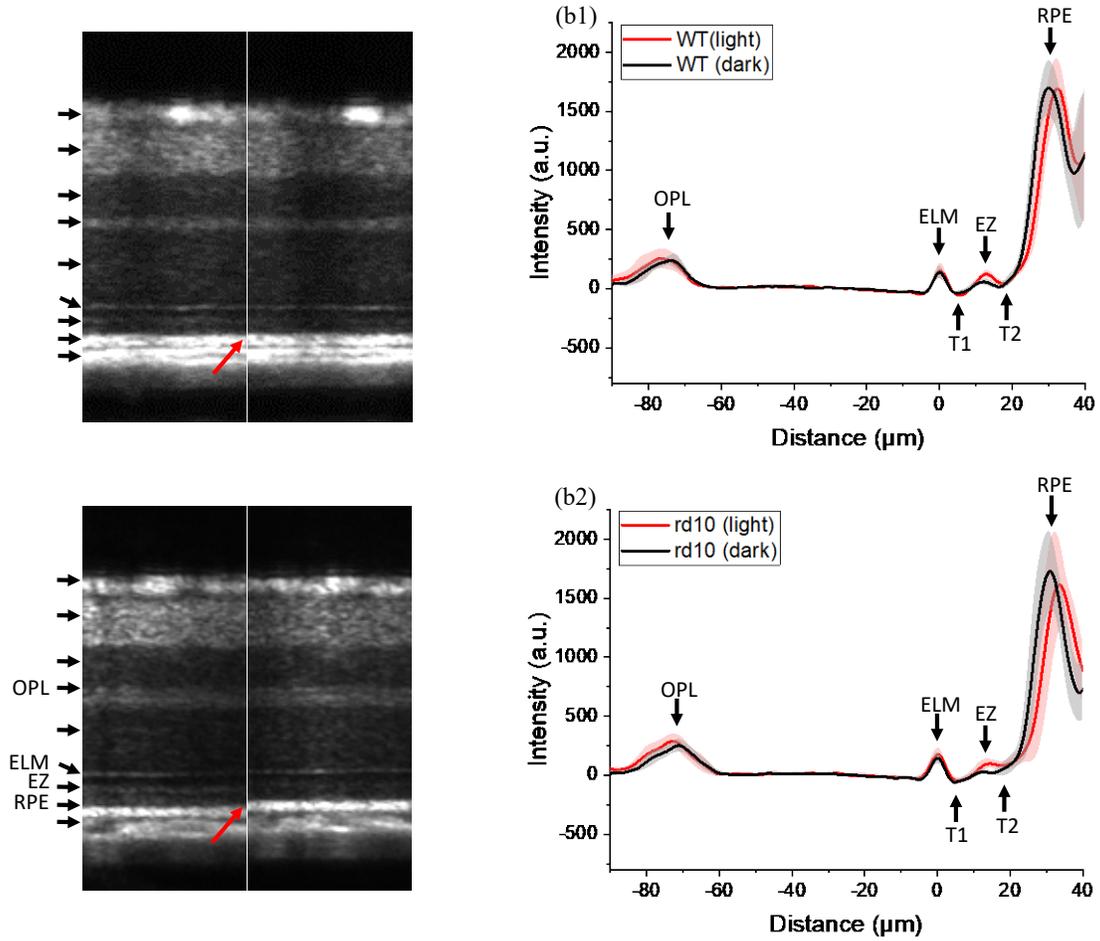

**Fig. 2** (a) Comparative OCT of WT (a1) and rd10 (a2) mouse retinas with LA and DA; (b) Average outer retina intensity profiles of WT (b1) and rd10 (b2) retinas with LA and DA. Standard deviations were included as corresponding shadows alongside the lines. NFL: nerve fiber layer; IPL: inner plexiform layer; INL: inner nuclear layer; OPL: outer plexiform layer; ONL: outer nuclear layer; ELM: external limiting membrane; EZ: inner segment ellipsoid; RPE: retinal pigment epithelium; Ch: choroid. WT sample size n = 6; rd10 sample size n = 8.

*3.2 Comparative Analysis of DA-IOS Kinetics of Outer Retina*

After confirming the outer retina thickness and OCT band intensity changes, we further compared DA-IOS kinetics in WT and rd10 mice at P14. Time-lapse OCT images were recorded from WT and rd10 mice during the 30 min DA. Seven OCT recordings were sequentially collected from 0 min (light) to 30 min with a 5 min interval for dynamic monitoring of DA-IOS kinetics.

First, the dynamic change of outer retinal thickness was quantitatively evaluated during the DA. Figure 3a illustrates the outer retina (OPL-RPE) thickness changes during DA in both WT and rd10 mice. The outer retinal thickness gradually reduced during DA, which is consistent with the observed outer retina shrinkage in DA condition in Fig. 2. For a better understanding of the thickness changes, Fig. 3b and Fig. 3c show separate analyses of the OPL-ELM and ELM-RPE thickness changes during DA. Both OPL-ELM and ELM-RPE thicknesses were consistently reduced during DA. The ELM-RPE thickness changes (WT: $\Delta$ELM-RPE = 2.0 μm; rd10: $\Delta$ELM-RPE = 2.5 μm, after 30 min DA) were significantly larger ($P<0.05$) than the OPL-ELM thickness changes (WT: $\Delta$OPL-ELM = 0.83 μm; rd10: $\Delta$OPL-ELM = 1.2 μm, after 30 min DA). For better comparative illustration, the WT and rd10 thicknesses were normalized to 1 in the light condition, i.e., at the time point 0 in Fig. 3d-3f. The relative thickness changes of OPL-ELM and OPL-RPE were similar in both group. However, the ELM-RPE thickness change in rd10 retina became significantly larger than WT after 20 min of DA (Fig. 3f).

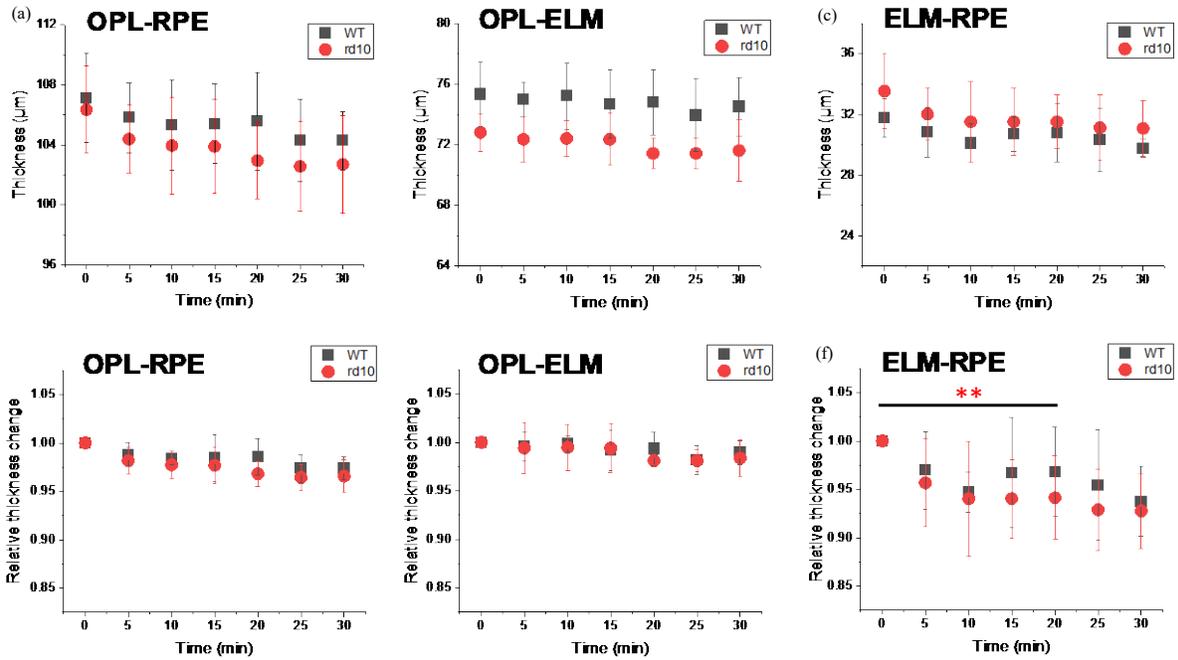

**Fig. 3** Comparative outer retina thickness analysis of WT and rd10 mice at P14. (a) Outer retina (OPL to RPE) thickness; (b) OPL to ELM thickness; (c) ELM to RPE thickness. (d) OPL-RPE relative thickness change; (e) OPL-ELM relative thickness change; (f) ELM-RPE relative thickness change; Standard deviation marked as corresponding bars alongside the lines. Only first *p*-value was marked. ** *p*<0.01, compared to light condition at 0 min, using one-way ANOVA with the Bonferroni correction. WT samples n = 6; rd10 samples n = 8.

In Fig. 2b, the OCT intensity of outer retina layers was observed to decrease in both groups. Relative intensity changes of outer retina layers during dark adaptation were then compared in Fig. 4. Results showed that relative intensity changes of OPL (Fig. 4a), ELM (Fig. 4b), T1 (Fig. 4d), and RPE (Fig. 4f) bands are not statistically different. However, DA-IOS kinetic difference was observed to be significant in EZ (Fig. 4c) and T2 (Fig. 4e). In the EZ, significant intensity reduction was observed in rd10, after 5 min DA (*p*<0.001), while it took 15 min to show significant intensity reduction (*p*<0.01) in WT (Fig.4c). This indicates that the EZ peak brightness decreased rapidly in rd10, compared to WT. Although the T2 intensity showed a significant decrease in rd10 after 5 min DA (*p*<0.001), the STD varies largely.

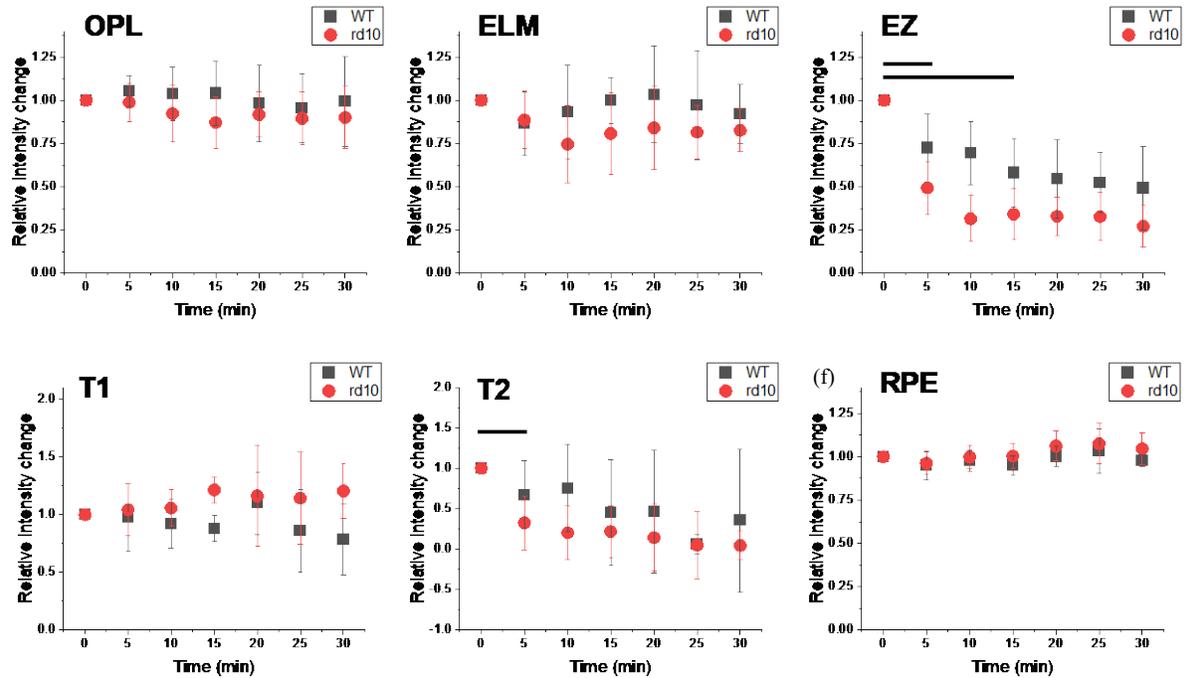

**Fig. 4** (a) OPL relative intensity change; (b) ELM relative intensity change; (c) EZ relative intensity change; (d) RPE relative intensity change. (e) T2 relative intensity change; (f) RPE relative intensity change. STD was marked as error bar. Only first *p*-value was marked. * $p<0.05$, ** $p<0.01$, *** $p<0.001$, compared to light condition at 0 min, using one-way ANOVA with the Bonferroni correction. WT sample size n = 6; rd10 sample size n = 8.

## 4 Discussion

In this study, functional OCT of rd10 mouse retina was conducted to assess abnormal retinal manifestations during DA. ELM-RPE thickness shrinkage and EZ intensity decrease were observed in both WT and rd10 retinas during DA (Fig. 3). With 30 min DA, the averaged thickness change of ELM-RPE in rd10 is larger than that in WT. Our results are consistent with previous studies, which have also reported outer retina shrinkage in dark condition[40,42,54]. As for the EZ intensity, it was observed to be significantly decreased in rd10 and WT retinas (Fig. 4). The difference is distinguishable in rd10 retina after 5 min in dark, while it was 15 min for WT retina to show significant EZ intensity decrease. These results suggest that DA-IOS kinetics can be a sensitive biomarker of rod degeneration.

Fast photoreceptor-IOS, which reflects the early phase of phototransduction, has been demonstrated in the retina[36]. Comparative study revealed similar fast photoreceptor-IOS in WT and rd10 mice at P14, confirming that the fast photoreceptor-IOS occurs before PDE activation in retinal photoreceptors[50]. Before P15, the relative expression level of rhodopsin and transducin in rd10 mice is known to be similar to that in WT mice[50,52,55]. In other words, the fast photoreceptor-IOS is not specifically sensitive to PDE deficiency in rd10 before P15. Thus, this experiment was designed to test if DA-IOS kinetics, which involves downstream of transducing activation, can be sensitive to PDE deficiency caused photoreceptor abnormality. Our results indicate that quantitative monitoring of DA kinetics provides supplemental information for functional assessment of photoreceptor physiology.

There are several possible mechanisms involved in the DA-IOS changes. Previous studies have suggested that enhanced RPE pumping activity of metabolic fluid[49,54,56,57] and morphological changes of mitochondria during DA[56,58] could result in reduced fluid content and reduction of reflective OCT band brightness, which contributes to the ELM-RPE thickness shrinkage and intensity decrease. It was also reported that oxidative stress of rod photoreceptor is higher in rd10[59], which may induce oxidative damage to photoreceptor cells and affect rod function[60]. The increasing supply of oxygen to photoreceptor region causes mitochondria to produce more respiratory metabolites including water content[61,62]. After DA, the accumulation of metabolic fluid in ELM-RPE is reduced and causes the thickness shrinkage in rd10 retina[54]. Therefore, we speculate that the mitochondrial activity with enhanced fluid removal is active in rd10 photoreceptor, leading to the accelerated ELM-RPE thickness shrinkage and EZ intensity reduction.

Another potential mechanism of the ELM-RPE thickness shrinkage may be that internal osmotic pressure was decreased in DA, leading to reversed cytoplasmic swelling with less backscattering. With light present, rhodopsin is activated, and transducin activates PDE for hydrolysis of cGMP, turning the ion channels off. This downstream of transducin activation may cause the osmotic pressure to increase with excess osmolytes, leading to cytoplasmic swelling and ELM-RPE thickness and EZ intensity increasing[63,64]. During DA with the absence of light, rhodopsin is inactivated, which makes the cascade unable to continue. As transducin no longer activates PDE, cGMP contributes to the ion flux through the opened $Na^+$ tunnel. Therefore, the internal osmotic pressure is decreased followed by cytoplasmic fluid removal, resulting in the ELM-RPE thickness shrinkage and EZ intensity reduction[63]. This phenomenon is downstream of transducin activation, of which the expression level in WT and rd10 at P14 is similar[50]. Thus, we observed changes of ELM-RPE thickness and EZ intensity in both groups. Although both groups had reductions of ELM-RPE thickness and EZ intensity, accelerated reductions were observed in rd10 at P14. This may result from the PDE deficiency in rd10 that is already existed at early development[47,50,52,55], causing the less functional PDE proteins with a higher accumulation of cGMP in the OS[46]. With the higher concentration of cGMP, the osmotic pressure may decrease quicker with restoration of cytoplasmic fluid removal during DA, and thus, result in shortening and less backscattering of the outer retina in rd10.

Interestingly, we also observed T2 intensity change in rd10 (Fig. 2b). The T2 relative intensity reduction was significantly decreased in rd10 retinas after 5 min DA, while T1 relative intensity change was not statistically significant. T1 was correlated with IS, while list T2 was correlated with OS. This observation suggests transient OS abnormality occurs during DA in rd10 retinas at P14. Further investigation is required to understand if this is related to the PDE deficiency and disruption of the OS region.

## 5 Conclusion

DA-IOS abnormalities were observed in rd10, compared to WT at P14. In dark conditions, WT and rd10 mice retinas were observed to show ELM-RPE thickness reduction and EZ intensity attenuation. The DA-associated thickness and intensity changes were larger in rd10, compared to WT. The hypo-reflective band T2 that corresponds to photoreceptor OS, shows abnormality in rd10 at P14. The ORG measurement of DA-IOS kinetics promises valuable information for noninvasive assessment of rod photoreceptor degeneration.

**Disclosures**

The authors declare that the research was conducted in the absence of any commercial or financial relationships that could be construed as a potential conflict of interest.

**Acknowledgments**

This research was supported in part by National Eye Institute (P30 EY001792, R01 EY023522, R01 EY030101, R01EY029673, R01EY030842); Research to Prevent Blindness; Richard and Loan Hill Endowment.

**Code, Data, and Materials Availability**

The data supporting the conclusions may be obtained from the authors upon reasonable request.

**References**

1. C. A. Curcio, N. E. Medeiros, and C. L. Millican, "Photoreceptor loss in age-related macular degeneration," *Invest Ophthalmol Vis Sci* **37**(7), 1236-1249 (1996).

2. S. Yang et al., "Photoreceptor dysfunction in early and intermediate age-related macular degeneration assessed with mfERG and spectral domain OCT," *Doc Ophthalmol* **132**(1), 17-26 (2016), doi:10.1007/s10633-016-9523-4.

3. G. R. Jackson, C. Owsley, and C. A. Curcio, "Photoreceptor degeneration and dysfunction in aging and age-related maculopathy," *Ageing Res Rev* **1**(3), 381-396 (2002).

4. E. L. Berson, P. Gouras, and R. D. Gunkel, "Rod responses in retinitis pigmentosa, dominantly inherited," *Arch Ophthalmol* **80**(1), 58-67 (1968), doi:10.1001/archopht.1968.00980050060009.

5. E. L. Berson, and E. B. Goldstein, "Early receptor potential in dominantly inherited retinitis pigmentosa," *Arch Ophthalmol* **83**(4), 412-420 (1970), doi:10.1001/archopht.1970.00990030412005.

6. K. Holopigian et al., "Evidence for photoreceptor changes in patients with diabetic retinopathy," *Invest Ophthalmol Vis Sci* **38**(11), 2355-2365 (1997).

7. J. J. McAnany, and J. C. Park, "Cone Photoreceptor Dysfunction in Early-Stage Diabetic Retinopathy: Association Between the Activation Phase of Cone Phototransduction and the Flicker Electroretinogram," *Invest Ophthalmol Vis Sci* **60**(1), 64-72 (2019), doi:10.1167/iovs.18-25946.

8. K. Ishikawa et al., "Focal macular electroretinograms after photodynamic therapy combined with intravitreal bevacizumab," *Graefes Archive for Clinical and Experimental Ophthalmology* **249**(2), 273-280 (2011), doi:DOI 10.1007/s00417-010-1548-x.

9.  B. Feigl, and C. P. Morris, "The challenge of predicting macular degeneration," *Curr Med Res Opin* **27**(9), 1745-1748 (2011), doi:10.1185/03007995.2011.603301.

10. J. S. Sunness et al., "Diminished foveal sensitivity may predict the development of advanced age-related macular degeneration," *Ophthalmology* **96**(3), 375-381 (1989).

11. D. Ts'o et al., "Noninvasive functional imaging of the retina reveals outer retinal and hemodynamic intrinsic optical signal origins," *Jpn J Ophthalmol* **53**(4), 334-344 (2009), doi:10.1007/s10384-009-0687-2.

12. X. Yao, and B. Wang, "Intrinsic optical signal imaging of retinal physiology: a review," *J Biomed Opt* **20**(9), 090901 (2015), doi:10.1117/1.JBO.20.9.090901.

13. G. Hanazono et al., "Evaluating neural activity of retinal ganglion cells by flash-evoked intrinsic signal imaging in macaque retina," *Invest Ophthalmol Vis Sci* **49**(10), 4655-4663 (2008), doi:10.1167/iovs.08-1936.

14. M. Begum, D. P. Joiner, and D. Y. Ts'o, "Stimulus-Driven Retinal Intrinsic Signal Optical Imaging in Mouse Demonstrates a Dominant Rod-Driven Component," *Invest Ophthalmol Vis Sci* **61**(8), 37 (2020), doi:10.1167/iovs.61.8.37.

15. T. Son et al., "Functional intrinsic optical signal imaging for objective optoretinography of human photoreceptors," *Exp Biol Med (Maywood)* **246**(6), 639-643 (2021), doi:10.1177/1535370220978898.

16. B. Wang, and X. Yao, "*In vivo* intrinsic optical signal imaging of mouse retinas," *Proc SPIE Int Soc Opt Eng* **9693**(2016), doi:10.1117/12.2212810.

17. Q. X. Zhang et al., "Comparative intrinsic optical signal imaging of wild-type and mutant mouse retinas," *Opt Express* **20**(7), 7646-7654 (2012), doi:10.1364/OE.20.007646.

18. X. C. Yao, "Intrinsic optical signal imaging of retinal activation," *Jpn J Ophthalmol* **53**(4), 327-333 (2009), doi:10.1007/s10384-009-0685-4.

19. X. C. Yao et al., "Rapid optical coherence tomography and recording functional scattering changes from activated frog retina," *Applied Optics* **44**(11), 2019-2023 (2005).

20. K. Bizheva et al., "Optophysiology: depth-resolved probing of retinal physiology with functional ultrahigh-resolution optical coherence tomography," *Proc Natl Acad Sci U S A* **103**(13), 5066-5071 (2006), doi:10.1073/pnas.0506997103.

21. M. Azimipour et al., "Optoretinogram: optical measurement of human cone and rod photoreceptor responses to light," *Opt Lett* **45**(17), 4658-4661 (2020), doi:10.1364/OL.398868.

22. V. P. Pandiyan et al., "The optoretinogram reveals the primary steps of phototransduction in the living human eye," *Sci Adv* **6**(37), (2020), doi:10.1126/sciadv.abc1124.

23. L. Zhang et al., "Volumetric data analysis enabled spatially resolved optoretinogram to measure the functional signals in the living retina," *J Biophotonics* 15(3):e202100252 (2021), doi:10.1002/jbio.202100252.

24. R. S. Jonnal, "Toward a clinical optoretinogram: a review of noninvasive, optical tests of retinal neural function," *Ann Transl Med* **9**(15), 1270 (2021), doi:10.21037/atm-20-6440.

25. P. Zhang et al., "Measurement of diurnal variation in rod outer segment length *in vivo* in Mice With the OCT Optoretinogram," *Invest Ophthalmol Vis Sci* **61**(3), 9 (2020), doi:10.1167/iovs.61.3.9.

26. A. Lassoued et al., "Cone photoreceptor dysfunction in retinitis pigmentosa revealed by optoretinography," *Proc Natl Acad Sci U S A* **118**(47), (2021), doi:10.1073/pnas.2107444118.

27. G. Ma et al., "Functional optoretinography: concurrent OCT monitoring of intrinsic signal amplitude and phase dynamics in human photoreceptors," *Biomed Opt Express* **12**(5), 2661-2669 (2021), doi:10.1364/BOE.423733.

28. R. F. Cooper, D. H. Brainard, and J. I. W. Morgan, "Optoretinography of individual human cone photoreceptors," *Opt Express* **28**(26), 39326-39339 (2020), doi:10.1364/OE.409193.

29. T. H. Kim et al., "Functional optical coherence tomography enables *in vivo* optoretinography of photoreceptor dysfunction due to retinal degeneration," *Biomed Opt Express* **11**(9), 5306-5320 (2020), doi:10.1364/BOE.399334.

30. V. P. Pandiyan et al., "High-speed adaptive optics line-scan OCT for cellular-resolution optoretinography," *Biomed Opt Express* **11**(9), 5274-5296 (2020), doi:10.1364/BOE.399034.

31. T.-H. Kim et al., "Functional Optical Coherence Tomography for Intrinsic Signal Optoretinography: Recent Developments and Deployment Challenges," Front. Med. **9**(864824) (2022), doi: 10.3389/fmed.2022.864824.

32. X. Yao et al., "Functional optical coherence tomography of retinal photoreceptors," *Exp Biol Med (Maywood)* **243**(17-18), 1256-1264 (2018), doi:10.1177/1535370218816517.

33. Q. Zhang et al., "Functional optical coherence tomography enables *in vivo* physiological assessment of retinal rod and cone photoreceptors," *Sci Rep* **5**(9595) (2015), doi:10.1038/srep09595.

34. B. Wang et al., "En face optical coherence tomography of transient light response at photoreceptor outer segments in living frog eyecup," *Opt Lett* **38**(22), 4526-4529 (2013), doi:10.1364/OL.38.004526.

35. A. A. Moayed et al., "*In vivo* imaging of intrinsic optical signals in chicken retina with functional optical coherence tomography," *Opt Lett* **36**(23), 4575-4577 (2011), doi:10.1364/OL.36.004575.

36. X. Yao, and T. H. Kim, "Fast intrinsic optical signal correlates with activation phase of phototransduction in retinal photoreceptors," *Exp Biol Med (Maywood)* **245**(13), 1087-1095 (2020), doi:10.1177/1535370220935406.

37. C. Owsley et al., "Delayed Rod-Mediated Dark Adaptation Is a Functional Biomarker for Incident Early Age-Related Macular Degeneration," *Ophthalmology* **123**(2), 344-351 (2016), doi:10.1016/j.ophtha.2015.09.041.

38. C. C. Hsiao et al., "Correlation of retinal vascular perfusion density with dark adaptation in diabetic retinopathy," *Graefes Arch Clin Exp Ophthalmol* **257**(7), 1401-1410 (2019), doi:10.1007/s00417-019-04321-2.

39. S. Gao et al., "Functional regulation of an outer retina hyporeflective band on optical coherence tomography images," *Sci Rep* **11**(1), 10260 (2021), doi:10.1038/s41598-021-89599-1.

40. Y. Li et al., "Light-Induced Thickening of Photoreceptor Outer Segment Layer Detected by Ultra-High Resolution OCT Imaging," *Invest Ophthalmol Vis Sci* **57**(9), OCT105-111 (2016), doi:10.1167/iovs.15-18539.

41. B. A. Berkowitz et al., "Mitochondrial respiration in outer retina contributes to light-evoked increase in hydration *in vivo*," *Invest Ophthalmol Vis Sci* **59**(15), 5957-5964 (2018), doi:10.1167/iovs.18-25682.

42. T. H. Kim, J. Ding, and X. Yao, "Intrinsic signal optoretinography of dark adaptation kinetics," *Sci Rep* **12**(1), 2475 (2022), doi:10.1038/s41598-022-06562-4.

43. B. A. Berkowitz et al., "Rod Photoreceptor Neuroprotection in Dark-Reared Pde6brd10 Mice," *Invest Ophthalmol Vis Sci* **61**(13), 14 (2020), doi:10.1167/iovs.61.13.14.

44. W. T. Deng et al., "Cone Phosphodiesterase-6gamma' Subunit Augments Cone PDE6 Holoenzyme Assembly and Stability in a Mouse Model Lacking Both Rod and Cone PDE6 Catalytic Subunits," *Front Mol Neurosci* **11**(233 (2018), doi:10.3389/fnmol.2018.00233.

45. S. Rosch et al., "Correlations between ERG, OCT, and Anatomical Findings in the rd10 Mouse," *J Ophthalmol* **2014**(874751) (2014), doi:10.1155/2014/874751.

46. T. Wang et al., "The PDE6 mutation in the rd10 retinal degeneration mouse model causes protein mislocalization and instability and promotes cell death through increased ion influx," *J Biol Chem* **293**(40), 15332-15346 (2018), doi:10.1074/jbc.RA118.004459.

47. M. E. Pennesi et al., "Long-term characterization of retinal degeneration in rd1 and rd10 mice using spectral domain optical coherence tomography," *Invest Ophthalmol Vis Sci* **53**(8), 4644-4656 (2012), doi:10.1167/iovs.12-9611.

48. N. L. H. B. Chang, R.E. Hurd, M.T. Davisson, S. Nusinowitz, J.R. Heckenlively, "Retinal degeneration mutants in the mouse," *Vision Research* **42**(517-525 (2002).

49. C. Gargini et al., "Retinal organization in the retinal degeneration 10 (rd10) mutant mouse: a morphological and ERG study," *J Comp Neurol* **500**(2), 222-238 (2007), doi:10.1002/cne.21144.

50. Y. Lu, T. H. Kim, and X. Yao, "Comparative study of wild-type and rd10 mice reveals transient intrinsic optical signal response before phosphodiesterase activation in retinal photoreceptors," *Exp Biol Med (Maywood)* **245**(4), 360-367 (2020), doi:10.1177/1535370219896284.

51. M. J. Phillips, D. C. Otteson, and D. M. Sherry, "Progression of neuronal and synaptic remodeling in the rd10 mouse model of retinitis pigmentosa," *J Comp Neurol* **518**(11), 2071-2089 (2010), doi:10.1002/cne.22322.

52. B. Chang et al., "Two mouse retinal degenerations caused by missense mutations in the beta-subunit of rod cGMP phosphodiesterase gene," *Vision Res* **47**(5), 624-633 (2007), doi:10.1016/j.visres.2006.11.020.

53. F. Mazzoni, H. Safa, and S. C. Finnemann, "Understanding photoreceptor outer segment phagocytosis: use and utility of RPE cells in culture," *Exp Eye Res* **126**, 51-60 (2014), doi:10.1016/j.exer.2014.01.010.

54. Y. Z. Yichao Li, Sonia Chen, Gregory Vernon, Wai T. Wong, and Haohua Qian, "Light-Dependent OCT Structure Changes in Photoreceptor Degenerative rd 10 Mouse Retina," *Invest Ophthalmol Vis Sci.* **59**, 1084-1094 (2018), doi:doi.org/ 10.1167/iovs.17-23011.

55. M. Samardzija et al., "Activation of survival pathways in the degenerating retina of rd10 mice," *Exp Eye Res* **99**, 17-26 (2012), doi:10.1016/j.exer.2012.04.004.

56. J. D. Linton et al., "Flow of energy in the outer retina in darkness and in light," *Proc Natl Acad Sci U S A* **107**(19), 8599-8604 (2010), doi:10.1073/pnas.1002471107.

57. D. B. B. A. Berkowitz, "Light-dependent changes in outer retinal water diffusion in rats *in vivo*," *Molecular Vision* 18), 2561-2577 (2012).

58. G. Ma et al., "*In vivo* optoretinography of phototransduction activation and energy metabolism in retinal photoreceptors," *J Biophotonics* **14**(5), e202000462 (2021), doi:10.1002/jbio.202000462.

59. B. A. Berkowitz et al., "Outer retinal oxidative stress measured *in vivo* using QUEnch-assiSTed (QUEST) OCT," *Invest Ophthalmol Vis Sci* **60**(5), 1566-1570 (2019), doi:10.1167/iovs.18-26164.

60. J. Wang et al., "Activation of the molecular chaperone, sigma 1 receptor, preserves cone function in a murine model of inherited retinal degeneration," *Proc Natl Acad Sci U S A* **113**(26), E3764-3772 (2016), doi:10.1073/pnas.1521749113.

61. L. Trachsel-Moncho et al., "Oxidative stress and autophagy-related changes during retinal degeneration and development," *Cell Death Dis* **9**(8), 812 (2018), doi:10.1038/s41419-018-0855-8.

62. C. Roehlecke et al., "Stress reaction in outer segments of photoreceptors after blue light irradiation," *PLoS One* **8**(9), e71570 (2013), doi:10.1371/journal.pone.0071570.

63. P. Zhang et al., "*In vivo* optophysiology reveals that G-protein activation triggers osmotic swelling and increased light scattering of rod photoreceptors," *Proc Natl Acad Sci U S A* **114**(14), E2937-E2946 (2017), doi:10.1073/pnas.1620572114.

64. C. D. Lu et al., "Photoreceptor layer thickness changes during dark adaptation observed with ultrahigh-resolution optical coherence tomography," *Invest Ophthalmol Vis Sci* **58**(11), 4632-4643 (2017), doi:10.1167/iovs.17-22171.

**Biographies**

**Jie Ding** is a Ph.D. student in Biomedical Engineering at the University of Illinois Chicago (UIC). She earned MS degree in Biology at the Illinois Institute of technology in 2019. After graduation, she worked at UIC for two years as research assistant and then joined Dr. Yao's lab for doctoral study in 2021. Her research focuses on optical imaging of retinal morphology and physiology.

**Tae-Hoon Kim** is a postdoctoral fellow at Genentech, Inc. He obtained his Ph.D. degree in Biomedical Engineering at the University of Illinois Chicago in 2022, and his BS and MS degree in Biomedical Engineering at the Catholic University of Daegu, South Korea. His research focuses on *in vivo* functional optical imaging of the retina.

**Guangying Ma** is a Ph.D. candidate in Biomedical Engineering at the University of Illinois Chicago. He obtained his MS and BS at Nankai University, China. Then he joined Dr. Yao's lab for doctoral study in 2018. He's research focuses on *in vivo* retinal imaging and biomedical image signal analysis.

**Xincheng Yao** is a Richard & Loan Hill Professor of Bioengineering and Ophthalmology & Visual Sciences at the University of Illinois at Chicago (UIC). Dr. Yao received his Ph.D. in Optics from the Institute of Physics, Chinese Academy of Sciences in 2001. This was followed by his postdoctoral research in the Biophysical Group at Los Alamos National Laboratory (LANL) from 2001 to 2004. He held a LANL Technical Staff appointment from 2004 to 2006 and served at CFD Research Corporation as a Senior Research Scientist from 2006 to 2007. He was a faculty of University of Alabama at Birmingham (UAB) from 2007-2014 and joined UIC faculty in 2014. Dr. Yao's research interests include biomedical optics instrumentation, ultra-widefield fundus photography, functional optical coherence tomography (OCT), OCT angiography, super-resolution ophthalmoscopy, and machine learning based image analysis and eye disease classification.

**Caption List**

**Fig. 1** (a) en face projection image of the dorsal quadrant in the mouse retina. (b) Representative averaged OCT B-scan of the mouse retina. Scale bar: 100 μm.

**Fig. 2** (a) Comparative OCT of WT (a1) and rd10 (a2) mouse retinas with LA and DA; (b) Average outer retina intensity profiles of WT (b1) and rd10 (b2) retinas under light and dark condition. Standard deviations were included as corresponding shadows alongside the lines. NFL: nerve fiber layer; IPL: inner plexiform layer; INL: inner nuclear layer; OPL: outer plexiform layer; ONL: outer nuclear layer; ELM: external limiting membrane; EZ: inner segment ellipsoid; RPE: retinal pigment epithelium; Ch: choroid. WT sample size n = 6; rd10 sample size n = 8.

**Fig. 3** Comparative outer retina thickness analysis of WT and rd10 mice at P14. (a) Outer retina (OPL to RPE) thickness; (b) OPL to ELM thickness; (c) ELM to RPE thickness. (d) OPL-RPE relative thickness change; (e) OPL-ELM relative thickness change; (f) ELM-RPE relative thickness change; Standard deviation marked as corresponding bars alongside the lines. Only first $p$-value was marked. ** $p<0.01$, compared to light condition at 0 min, using one-way ANOVA with the Bonferroni correction. WT samples n = 6; rd10 samples n = 8.

**Fig. 4** (a) OPL relative intensity change; (b) ELM relative intensity change; (c) EZ relative intensity change; (d) RPE relative intensity change. (e) T2 relative intensity change; (f) RPE relative intensity change. STD was marked as error bar. Only first $p$-value was marked. * $p<0.05$, ** $p<0.01$, *** $p<0.001$, compared to light condition at 0 min, using one-way ANOVA with the Bonferroni correction. WT sample size n = 6; rd10 sample size n = 8.